\documentclass[hyper]{JHEP} % 10pt is ignored!
\newcommand\fverb{\setbox\pippobox=\hbox\bgroup\verb}
\newcommand\fverbdo{\egroup\medskip\noindent%
            \fbox{\unhbox\pippobox}\ }

\newcommand\fverbit{\egroup\item[\fbox{\unhbox\pippobox}]}
\newbox\pippobox
\title{Spiky Strings on I-brane}
\author{Sagar Biswas\\
Department of Phyisics and Meteorology, \\
Indian Institute of technology Kharagpur,\\
Kharagpur-721 302, INDIA \\
\email{biswas.sagar09iitkgp@gmail.com}}
\author{Kamal L. Panigrahi\\
Department of Physics and Meteorology, \\
Indian Institute of Technology Kharagpur,\\
Kharagpur-721302, INDIA, \\
and  \\
The Abdus Salam International Centre for Theoretical Physics, \\
Strada Costiera 11, Trieste, ITALY \\
E-mail: \email{panigrahi@phy.iitkgp.ernet.in}} \abstract{We study
rigidly rotating strings in the near horizon geometry of the 1+1
dimensional intersection of two orthogonal stacks of NS5-branes,
the so called I-brane background. We solve the equations of motion
of the fundamental string action in the presence of two form
NS-NS fluxes that the I-brane background supports and
write down general form of conserved quantities. We further find
out two limiting cases corresponding to giant magnon and single
spike like strings in various parameter space of solutions.}

\begin{document}
%%%%%%%%%%%%%%%%%%%%%%%%%%%%%%%%%%
%%%%%%%%%%%%%%%%%%%%%%%%%%%%%%%%%%
\section{Introduction}
The AdS/CFT duality \cite{Maldacena:1997re} relates the spectrum of free strings in the
bulk to that of field theory operators living on the boundary of
AdS. Though finding the full string theory spectrum is extremely
difficult it has been observed that in certain limits, for
example, in large angular momentum limit the theory becomes
tractable \cite{Berenstein:2002jq} and it is interesting to study the corresponding field
theory in detail. It has been noticed that in this region one can use the
semiclassical approximation to find the string spectrum as well \cite{Gubser:2002tv}.
Further an interesting observation is that the ${\cal N} =4$ SYM in planar limit can
be described by an integrable spin chain model where the
anomalous dimension of the gauge invariant operators were found in
\cite{Beisert:2005tm, Minahan:2002ve, Beisert:2003xu, Beisert:2003yb, Beisert:2004hm}, and it
was also noticed that the string theory is integrable in the semiclassical limit, see for example
\cite{Pohlmeyer:1975nb, Tseytlin:2004xa, Hayashi:2007bq, Okamura:2008jm}.

Study of rigidly rotating string in
semiclassical approximation has been one of the interesting areas
of research in the last few years because of its elegance. In this
connection the so called Hofman-Maldacena (HM) limit\footnote{The HM limit:
$J\rightarrow \infty, ~ \lambda = {\rm fixed}, ~ p= {\rm fixed}, ~ E-J = {\rm fixed}$, where $J$ is
one of the SO(6) charges and $p$ is the magnon momentum.} \cite{Hofman:2006xt} simplifies
considerably the problem of finding out the spectrum on both sides
of the duality. The spectrum consists of an elementary excitation
known as magnon which propagates with a conserved momentum $p$
along the spin chain. Further, a more general class of rotating
string solution in $AdS_5$ is the spiky string which describes the
higher twist operators from dual field theory view point \cite{Kruczenski:2004wg} and
magnon solutions can be thought of as a subspace of these spike
solutions. Infact it was further determined in \cite{Ishizeki:2007we} that if
one solves the most general form of equations of motion for a
rigidly rotating string on a sphere one encounters two set of
solutions, corresponding precisely to the giant magnon and single
spike solutions.

To understand the AdS/CFT like dualities in  more general backgrounds, it
is instructive to study rigidly rotating string in the gravity side which will
teach us the corresponding operators in the dual field theory side. In this
connection, magnon like dispersion relations have been found out in
backgrounds which arise from intersecting branes in supergravities
and which asymptotes to AdS and non-AdS backgrounds, see for example \cite{Bobev:2005cz, Ryang:2005pg,
Chu:2006ae, Kruczenski:2006pk, Kluson:2007qu, Ishizeki:2007we,
Bobev:2007bm, Dimov:2007ey, Kluson:2007fr, Lee:2008sk, Bykov:2008bj, David:2008yk, Kluson:2008gf, Lee:2008ui,
Ryang:2008rc, Abbott:2008qd, Suzuki:2009sc, Abbott:2009um, Biswas:2011wu}.
One of such background is the so-called I-brane
background \cite{Itzhaki:2005tu} that arises in the 1+1 dimensional intersection of two orthogonal stacks of NS5-branes,
with one set of branes lying along $(x^0, x^1, \cdots , x^5)$, and other set of lying along
$(x^0, x^1, x^6, \cdots , x^9)$ directions.
When all five branes are coincident, in the S-dual picture, the near horizon geometry is given by
\begin{eqnarray}
{\rm R}^{2,1} \times {\rm R}_{\phi} \times {\rm SU}(2)_{{k_1}} \times {\rm SU}(2)_{{k_2}} \ ,
\nonumber
\end{eqnarray}
where, $R_{\phi}$ is one combination of the radial directions away from the two
sets of NS5-branes, and the coordinates of $R^{2,1}$ are $x^0, x^1$ and
another combination of the two radial directions.
The two $SU(2)$'s describe the angular three-spheres corresponding to $({\bf R^4})_{2345}$ and
$({\bf R^4})_{6789}$.
As mentioned in \cite{Itzhaki:2005tu} this geometry has the interesting property that it exhibits
a higher symmetry than the full brane configuration itself. In particular, the combination of
radial directions away from the intersection that enters in ${\rm R}^{2,1}$ appears symmetrically with
the other spatial direction, and the background has a higher Poincare symmetry, ISO(2, 1),
than the expected ISO(1, 1) and twice as many supercharges one might expect.
Furthermore, in \cite{Kluson:2005eb, Kluson:2005qq, Hung:2006nn, Kluson:2007st},
these interesting properties were studied from the
point of view of D1-brane probe. It was shown that the enhancement of the near horizon
geometry has clear impact on the worldvolume dynamics of the D1-brane probe dynamics.

In this paper, we would like to study the solutions of rigidly rotating strings in the near horizon
geometry of this interesting background with an emphasis on finding out spiky strings. Earlier
in \cite{Kluson:2008gf} a class of such rigidly rotating strings were studied and some interesting solutions
were obtained. In this paper we wish to find out solutions corresponding to the usual giant
magnon and single spike solutions on $R_t \times S^3$ and $R \times S^3 \times S^3$ in the presence of
various background NS-NS fluxes that it supports. Even though we can not compare our results to
the dual field theory which is not completely known, but a knowledge of the bulk solutions gives us
idea weather to look for operators in dual theory. The rest of the paper is organized as follows. In section-2
we give a brief overview of I-brane background and write down the near horizon geometry in a parametrization
which we will
use in the subsequent analysis. In section-3 we write the Nambu-Goto action for the Fundamental
string in the near horizon geometry of I-brane, write down the conserved charges and compute the
general equations of motion of the rigidly rotating string. Further we find out, in some parameter space of
solution, the two limiting cases corresponding to giant magnon and single spike like strings corresponding to
open string boundary condition. Finally in section-4 we conclude with some remarks.

\section{Review of near horizon geometry of I-brane} In this section we would like to
give a brief review of the near horizon geometry of the I-brane which arises from the
1+1 dimensional intersection of two orthogonal stacks of NS5-branes. This background arises
when $k_1$ number of NS5-branes lying along $(0, 1, \cdots , 5)$ intersect $k_2$ number
of NS5-branes lying along $(0, 1, 6, \cdots , 9)$ directions in $(0, 1)$-plane.
If the branes are coincident, then the supergravity solution is given by \cite{Itzhaki:2005tu}
\begin{eqnarray}
ds^2 &=& -{(dx^0)}^2 + {(dx^1)}^2 + H_{1} (y) \sum^5_{\alpha = 2} {(dy^{\alpha})}^2 +
H_{2} (z) \sum^9_{p = 6} {(dz^p)}^2 , \>\>\> e^{2\phi} = H_{1} (y) H_{2} (z) , \nonumber \\
H_{\alpha\beta\gamma} &=& -\epsilon_{\alpha\beta\gamma\delta}\partial^{\delta} H_1 (y) , \>\>
H_{mnp} = -\epsilon_{mnpq}\partial^q H_2 (z) ,\>\>  H_1 = 1 + \frac{k_1{l_s}^2}{y^2}, \>\>   H_2 = 1+ \frac{k_2{l_s}^2}{z^2}  \ ,\nonumber \\
y &=& \sqrt{\sum^{5}_{\alpha =2} {(y^{\alpha})}^2}, \>\>\>\>
z = \sqrt{\sum^{9}_{p=6} {(z^{p})}^2} \ .
\end{eqnarray}
In the near horizon limit the 1 in the Harmonic functions above can be neglected and the
metric and NS-NS ($B_{\mu\nu}$) fields are given by \footnote{This is
also supported by a dilaton whose explicit form we will not need in the calculation of this paper.}
\begin{eqnarray}
ds^2 &=& -{(dx^0)}^2 +  {(dx^1)}^2 + k_1 l^2_s \frac{dr_1^2}{r_1^2}+ k_1 l^2_s d\Omega_1^2
+ k_2 l^2_s \frac{dr_2^2}{r_2^2}+ k_2 l^2_s d\Omega_2^2 \, \nonumber \\
B_{\phi_1\psi_1} &=& 2k_1 l^2_s \sin^2\theta_1  \ , \>\>\>\> B_{\phi_2\psi_2} = 2k_2 l^2_s \sin^2\theta_2 \ ,
%\nonumber \\
%dy^2 &=& dr^2_1 + r^2_1 d\Omega^2_1 \ ,\>\>\>\>  dz^2 = dr^2_2 + r^2_2 d\Omega^2_2
\end{eqnarray}
where
%$dy^2 = dr^2_1 + r^2_1 d\Omega^2_1,  dz^2 = dr^2_2 + r^2_2 d\Omega^2_2$, and
$$d \Omega^2_1 = d{\theta_1}^2 + \sin^2\theta_1d{\phi_1}^2 + \cos^2\theta_1d{\psi_1}^2 \ , \>\>\>\>
d \Omega^2_2= d{\theta_2}^2 + \sin^2\theta_2d{\phi_2}^2 +
\cos^2\theta_2d{\psi_2}^2 .$$ are the volume elements on the sphere along $(y^2, \cdots, y^5)$
and $(z^6, \cdots, z^9)$ directions respectively.
%To proceed further we will
%set $k_1 = k_2 = N$ (the same number of NS5-branes in the two different stacks) \footnote{from now on we %will also work in the unit $l_s =1$.} and introduce the following coordinates to describes the
%spheres
%\begin{eqnarray}
%x^2 + ix^3 &=& r_1\cos\theta_1e^{i\phi_1}, \>\>  x^4 + ix^5 = r_1\cos\theta_1e^{i\psi_1}, \nonumber \\
%x^6 + ix^7 &=& r_2\cos\theta_2e^{i\phi_2}, \>\>  x^8 + ix^9 = r_2\cos\theta_2e^{i\psi_2},
%\end{eqnarray}
%so that the volume elements on the sphere, and the 2-form potentials which give the
%volume forms in eqn. (\ref{I-bkg}) are given by
%\begin{eqnarray}
%d{\Omega_1}^2 &=& d{\theta_1}^2 + \sin^2\theta_1d{\phi_1}^2 +
%\cos^2\theta_1d{\psi_1}^2 \ , \>\>\>\> B_{\phi_1\psi_1} = 2N \sin^2\theta_1 \ , \nonumber \\
%\>\> 0 < \theta_1 < \frac{\pi}{2}, \>\>
%0 = \phi_1,\psi_1 < 2\pi, \nonumber \\
%d{\Omega_2}^2&=& d{\theta_2}^2 + \sin^2\theta_2d{\phi_2}^2 +
%\cos^2\theta_2d{\psi_2}^2 \ , \>\>\>\> B_{\phi_2\psi_2} = 2N \sin^2\theta_2 \ .
%\>\> 0 < \theta_2 < \frac{\pi}{2}, \>\>
%0 = \phi_2,\psi_2 < 2\pi,
%\end{eqnarray}
To proceed further we make the following change of variables (we choose $l_s = 1, k_1 = k_2 = N$)
\begin{eqnarray}
\rho_1=\ln \frac{r_1}{\sqrt{N}}, \>\>\>\> \rho_2=\ln \frac{r_2}{\sqrt{N}},\>\>\>
x^0 = \sqrt{N} t, \>\>\>\>\> x^1 = \sqrt{N} y \ .
\end{eqnarray}
The final form of metric and the background NS-NS fields are given by
\begin{eqnarray}
ds^2 &=& N(-dt^2 + dy^2 + d\rho_1^2 + d\theta_1^2 + \sin^2\theta_1 d\phi_1^2 +\cos^2\theta_1 d\psi_1^2
+ d\rho_2^2 + d\theta_2^2 + \nonumber \\
&+& \sin^2\theta_2 d\phi_2^2 +\cos^2\theta_2 d\psi_2^2) \ , \>\>\>
B_{\phi_1 \psi_1} = 2N \sin^2\theta_1, \>\>\>\> B_{\phi_2 \psi_2} = 2N \sin^2\theta_2 . \nonumber \\
\label{I-brane}
\end{eqnarray}
We are interested in studying the rigidly rotating strings and finding out spiky strings in this background.\footnote{ Recently in \cite{Nayak:2010bw} a PP-wave background has been obtained by applying Penrose limit on this geometry.}
\section{Rigidly rotating strings in I-brane background} In this section we would like
to study the solutions to the fundamental string equations of motion in the background
discussed in the previous section. The Nambu-Goto action for the string in the background of
(\ref{I-brane}) is written as
\begin{eqnarray}
S &=& -\frac{\sqrt{\lambda}}{2\pi} \int d^2\sigma \mathcal{L}, \,
\end{eqnarray}
where the Lagrangian $\mathcal{L}$ is given by
\begin{eqnarray}
\mathcal{L} &=& \Big(\Big[\Big(-\dot{t} t^{\prime} + \dot{y} y^{\prime} +
\dot{\rho_1}\rho_1^{\prime} + \dot{\rho_2}\rho_2^{\prime} +
\dot{\theta_1}\theta_1^{\prime}+ \dot{\theta_2}\theta_2^{\prime}
+ \sin^2\theta_1 \dot{\phi_1} \phi_1^{\prime} + \sin^2\theta_2 \dot{\phi_2} \phi_2^{\prime} \nonumber \\
&+& \cos^2\theta_1 \dot{\psi_1} \psi_1^{\prime} + \cos^2\theta_2 \dot{\psi_1} \psi_2^{\prime}\Big)^2
- \Big(-{\dot t}^2 + \dot{y}^2 + {\dot{\rho}^2}_1 + {\dot{\rho}^2}_2 + {\dot{\theta}^2}_1 + {\dot{\theta}^2}_2 +
\sin^2\theta_1 {\dot{\phi^2}}_1 \nonumber \\
&+& \sin^2\theta_2 {\dot{\phi^2}}_2
+ \cos^2\theta_1 {\dot{\psi^2}}_1 + \cos^2\theta_2 {\dot{\psi^2}}_2\Big)\Big( -{t^{\prime}}^2
+ {y^{\prime}}^2 + {\rho_1^{\prime}}^2 + {\rho_2^{\prime}}^2 + {\theta_1^{\prime}}^2 +
{\theta_2^{\prime}}^2 \nonumber \\
&+& \sin^2 \theta_1 {\phi_1^{\prime}}^2 + \sin^2 \theta_2 {\phi_1^{\prime}}^2
+ \cos^2 \theta_1 {\psi_1^{\prime}}^2 + \cos^2 \theta_1 {\psi_2^{\prime}}^2\Big)\Big]^{1/2} \nonumber \\
&+& 2\sin^2\theta_1(\dot{\phi}_1 \psi_1^{\prime} - \dot{\psi}_1
\phi_1^{\prime}) + 2\sin^2\theta_2(\dot{\phi}_2 \psi_2^{\prime} - \dot{\psi}_2 \phi_2^{\prime})\Big) \ ,
\end{eqnarray}
%\begin{eqnarray}
%[\sqrt{-{\rm det }g} + \epsilon^{\alpha \beta} B_{\mu \nu}
%\partial_{\alpha}X^{\mu} \partial_{\beta}X^{\nu}] \\ \nonumber &=& -\frac{\sqrt{\lambda}}{2\pi}
%\int d^2 \sigma \mathcal{L} \ ,
%\end{eqnarray}
%where the Lagrangian density is given by
%\begin{equation}
%\mathcal{L} = \sqrt{-det {\rm g}} + 2\sin^2\theta_1(\dot{\phi}_1 \psi_1^{\prime} - \dot{\psi}_1
%\phi_1^{\prime}) + 2\sin^2\theta_2(\dot{\phi}_2 \psi_2^{\prime} - \dot{\psi}_2 \phi_2^{\prime}) \ ,
%\end{equation}
%with $g_{\alpha \beta} = \partial_{\alpha} X^{\mu} \partial_{\beta} X^{\nu} G_{\mu\nu}$
%being the pull back of the spacetime metric onto the worldsheet
and $\sqrt{\lambda} = N$ is the 't Hooft coupling constant.
For studying the rigidly rotating strings we choose the following ansatz
\begin{eqnarray}
t&=&\kappa \tau,\>\> y = v\tau , \>\> \> \rho_i=m_i\tau, \>\>\> \theta_i=\theta_i(\sigma),\>\>\> \phi_i=\nu_i\tau +
\sigma,\>\>\> \psi_i = \omega_i \tau + \psi_i(\sigma), \>\>\> i = 1, 2 \ , \nonumber \\
%y &=& v\tau \>\>\> \rho_2=m_2\tau,
%\>\>\> \theta_2=\theta_2(\sigma), \>\>\> \phi_2=\nu_2\tau + \sigma, \>\>\> \psi_2= \omega_2 \tau +
%\psi_2(\sigma).
\end{eqnarray}
The Euler-Lagrangian equations derived from the above action are given by,
\begin{equation}
\partial_{\sigma} \frac{\partial \mathcal{L}}{\partial t^{\prime}} + \partial_{\tau}
\frac{\partial \mathcal{L}}{\partial \dot{t}} = \frac{\partial \mathcal{L}}{\partial t} \ ,
\end{equation}
\begin{equation}
\partial_{\sigma} \frac{\partial \mathcal{L}}{\partial y^{\prime}} + \partial_{\tau}
\frac{\partial \mathcal{L}}{\partial \dot{y}} = \frac{\partial \mathcal{L}}{\partial y} \ ,
\end{equation}
\begin{equation}
\partial_{\sigma} \frac{\partial \mathcal{L}}{\partial \rho_i^{\prime}}  + \partial_{\tau}
\frac{\partial \mathcal{L}}{\partial \dot{\rho}_i} = \frac{\partial \mathcal{L}}{\partial \rho_i} \ ,
\end{equation}
%\begin{equation}
%\partial_{\sigma} \frac{\partial \mathcal{L}}{\partial \rho_2^{\prime}} + \partial_{\tau}
%\frac{\partial \mathcal{L}}{\partial \dot{\rho}_2} = \frac{\partial \mathcal{L}}{\partial \rho_2}
%\end{equation}
\begin{equation}
\partial_{\sigma} \frac{\partial \mathcal{L}}{\partial \phi_i^{\prime}}  + \partial_{\tau}
\frac{\partial \mathcal{L}}{\partial \dot{\phi}_i} = \frac{\partial \mathcal{L}}{\partial \phi_i} \ ,
\end{equation}
%\begin{equation}
%\partial_{\sigma} \frac{\partial \mathcal{L}}{\partial \phi_2^{\prime}} + \partial_{\tau}
%\frac{\partial \mathcal{L}}{\partial \dot{\phi}_2} = \frac{\partial \mathcal{L}}{\partial \phi_2}
%\end{equation}
\begin{equation}
\partial_{\sigma} \frac{\partial \mathcal{L}}{\partial \psi_i^{\prime}}  + \partial_{\tau}
\frac{\partial \mathcal{L}}{\partial \dot{\psi}_i} = \frac{\partial \mathcal{L}}{\partial \psi_i} \ .
\end{equation}
%\begin{equation}
%\partial_{\sigma} \frac{\partial \mathcal{L}}{\partial \psi_2^{\prime}} + \partial_{\tau}
%\frac{\partial \mathcal{L}}{\partial \dot{\psi}_2} = \frac{\partial \mathcal{L}}{\partial \psi_2}
%\end{equation}
Next we have to solve these equations by the ansatz that we have proposed. Solving for
$t, \psi_1$ and $\psi_2$ we get the following
\begin{eqnarray}
[C_1 \nu_1^2 \sin^2 \theta_1 + C_1 \nu_2^2 \sin^2 \theta_2 + C_1 \omega_2^2 \cos^2 \theta_2 + 2\kappa
\nu_1 \omega_1 \sin^2 \theta_1 -\alpha^2 C_1 - \kappa \omega_1 C_4] \psi_1^{\prime} \cos^2 \theta_1
\nonumber \\
=[\kappa C_4 - 2\kappa \nu_1 \sin^2 \theta_1 + C_1\omega_1 \cos^2\theta_1][\omega_2 \psi_2^{\prime}
\cos^2 \theta_2 + \nu_1 \sin^2 \theta_1 + \nu_2 \sin^2 \theta_2] , \nonumber \\
\label{1st}
\end{eqnarray}
\begin{eqnarray}
[C_1 \nu_1^2 \sin^2 \theta_1 + C_1 \nu_2^2 \sin^2 \theta_2 + C_1 \omega_1^2 \cos^2 \theta_1 + 2\kappa
\nu_2 \omega_2 \sin^2 \theta_2 -\alpha^2 C_1 - \kappa \omega_2 C_5] \psi_2^{\prime} \cos^2 \theta_2
\nonumber \\
=[\kappa C_5 - 2\kappa \nu_2 \sin^2 \theta_2 + C_1\omega_2 \cos^2\theta_2][\omega_1 \psi_1^{\prime}
\cos^2 \theta_1 + \nu_1 \sin^2 \theta_1 + \nu_2 \sin^2 \theta_2] , \nonumber \\
\label{2nd}
\end{eqnarray}
and
\begin{eqnarray}
{\theta_1^{\prime}}^2 + {\theta_2^{\prime}}^2 = \frac{\kappa^2 - C_1^2}{C_1^2} \frac{(\nu_1 \sin^2
\theta_1 + \omega_1 \psi_1^{\prime} \cos^2\theta_1 + \nu_2 \sin^2 \theta_2
+ \omega_2 \psi_2^{\prime} \cos^2\theta_2)^2}{(\alpha^2 - \nu_1^2\sin^2\theta_1 -
\nu_2^2\sin^2\theta_2 - \omega_1^2\cos^2\theta_1 - \omega_2^2\cos^2\theta_2)} \nonumber \\
- \sin^2 \theta_1 - \sin^2 \theta_2 - {\psi_1^{\prime}}^2 \cos^2 \theta_1
- {\psi_2^{\prime}}^2 \cos^2  \theta_2 , \nonumber \\
\label{theta}
\end{eqnarray}
where $C_1$, $C_4$ and $C_5$ are integration constants and $\alpha = \sqrt{\kappa^2 - v^2 -m_1^2
-m_2^2}$. Note also that the Lagrangian is invariant under the following transformation of the fields,
\begin{eqnarray}
t^{\prime}(\tau,\sigma)=t(\tau,\sigma) + \epsilon_t,\>\>\> y^{\prime}(\tau,\sigma)=y(\tau,\sigma) +
\epsilon_y,\>\>\> \rho_1^{\prime}(\tau,\sigma)=\rho_1(\tau,\sigma) + \epsilon_{\rho_1}\nonumber \\
\rho_2^{\prime}(\tau,\sigma)=\rho_2(\tau,\sigma) + \epsilon_{\rho_2}, \>\>\>\>
\phi_1^{\prime}(\tau,\sigma)=\phi_1(\tau,\sigma) + \epsilon_{\phi_1}, \>\>\>
\phi_2^{\prime}(\tau,\sigma)=\phi_2(\tau,\sigma) + \epsilon_{\phi_2}, \nonumber \\
\psi_1^{\prime}(\tau,\sigma)=\psi_1(\tau,\sigma) + \epsilon_{\psi_1},\>\>\>
\psi_2^{\prime}(\tau,\sigma)=\psi_2(\tau,\sigma) + \epsilon_{\psi_2}, \nonumber \\
\end{eqnarray}
where $\epsilon_t$, $\epsilon_y$, $\epsilon_{\rho_1}$, $\epsilon_{\rho_2}$, $\epsilon_{\phi_1}$,
$\epsilon_{\phi_2}$, $\epsilon_{\psi_1}$, $\epsilon_{\psi_2}$ are constants. Hence the rigidly
rotating string has a number of conserved charges, namely the total energy $E$, angular momenta
$J_1$, $J_2$, $K_1$ and $K_2$ corresponding to the translation along $\phi_1$, $\phi_2$,
$\psi_1$ and $\psi_2$ directions respectively. The Conserved charge $P_y$, $D_1$ and $D_2$ are due to
the translational invariance of $y$, $\rho_1$ and $\rho_2$ coordinates respectively.
Now looking at (\ref{theta}), it is very hard to solve the equation in complete generality.
Below we will look for solutions in some limits of $\theta_1$ and $\theta_2$.

\subsection{${\rm For}~\theta_2= 0$ and $\theta_1=\theta$}
In this section we wish to study the case when the string is constrained to
rotate only one one sphere, effectively making the string dynamics on $R_t \times S^3$. In this case the equation of motion simplifies considerably. We have to essentially solve the following two equations
\begin{eqnarray}
[C_1 \nu_1^2 \sin^2 \theta  + C_1 \omega_2^2  + 2\kappa
\nu_1 \omega_1 \sin^2 \theta -\alpha^2 C_1 - \kappa \omega_1 C_4] \psi_1^{\prime} \cos^2 \theta
\nonumber \\
=[\kappa C_4 - 2\kappa \nu_1 \sin^2 \theta + C_1\omega_1 \cos^2\theta][\omega_2 \psi_2^{\prime} + \nu_1 \sin^2 \theta] , \nonumber \\
\label{1st1}
\end{eqnarray}
\begin{eqnarray}
[C_1 \nu_1^2 \sin^2 \theta  + C_1 \omega_1^2 \cos^2 \theta -\alpha^2 C_1 - \kappa \omega_2 C_5] \psi_2^{\prime}
=[\kappa C_5  + C_1\omega_2][\omega_1 \psi_1^{\prime}
\cos^2 \theta + \nu_1 \sin^2 \theta] , \nonumber \\
\label{2nd1}
\end{eqnarray}
To make things look simpler, let us define the following constants
\begin{eqnarray}
a=C_1\nu_1^2\sin^2\theta - C_1 \alpha^2,\>\>\>
b=C_1\omega_2^2  + 2\kappa \nu_1 \omega_1 \sin^2\theta - \kappa \omega_1 C_4,
c=C_1\omega_1^2 \cos^2\theta  - \kappa \omega_2 C_5, \nonumber \\
d=\nu_1\sin^2 \theta ,\>\>\>
e= \kappa C_4 -2\kappa \nu_1\sin^2\theta + C_1 \omega_1\cos^2 \theta, \>\>\>
f= \kappa C_5  + C_1 \omega_2 . \nonumber \\
\end{eqnarray}
With the above identifications, (\ref{1st}) and (\ref{2nd}) reduce to
\begin{eqnarray}
(a + b)\psi_1^{\prime}\cos^2 \theta = e(\omega_2 \psi_2^{\prime} + d) \ , \nonumber \\
(a + c)\psi_2^{\prime} = f(\omega_1 \psi_1^{\prime}\cos^2 \theta + d) \ .
\nonumber
\end{eqnarray}
From these we can easily solve for $\psi_1^{\prime}$ and $\psi_2^{\prime}$, and they are given by,
\begin{equation}
\psi_1^{\prime} = \frac{de(a + c + \omega_2 f)}{\cos^2\theta[(a+b)(a+c) - \omega_1 \omega_2 ef]} \ ,
\end{equation}
\begin{equation}
\psi_2^{\prime} = \frac{df(a + b + \omega_1 e)}{[(a+b)(a+c) - \omega_1 \omega_2 ef]} \ .
\end{equation}
Further (\ref{theta}) is now given by
\begin{eqnarray}
{\theta^{\prime}}^2  = \frac{\kappa^2 - C_1^2}{C_1^2} \frac{(\nu_1 \sin^2
\theta + \omega_1 \psi_1^{\prime} \cos^2\theta
+ \omega_2 \psi_2^{\prime} )}{(\alpha^2 - \nu_1^2\sin^2\theta  - \omega_1^2\cos^2\theta - \omega_2^2)} - \sin^2 \theta  - {\psi_1^{\prime}}^2 \cos^2 \theta
- {\psi_2^{\prime}}^2 \ . \nonumber \\
\label{theta1}
\end{eqnarray}
Below we will discuss two limiting cases that correspond to single spike and giant magnon solutions.
\subsubsection{Single Spike solution}
In the limit $\theta^{\prime} \rightarrow 0$ as $\theta \rightarrow \frac{\pi}{2}$, we obtain the
condition
\begin{eqnarray}
C_1= \frac{\kappa \nu_1}{\sqrt{\alpha^2 - \omega_2^2}} \ ,
\end{eqnarray}
and
$\psi_1^{\prime} \rightarrow 0$  as $\theta \rightarrow
\frac{\pi}{2}$ implies $C_4=2\nu_1$ also $\psi_2^{\prime} \rightarrow 0$  as $\theta \rightarrow
\frac{\pi}{2}$ implies $C_5=-\frac{C_1 \omega_2}{\kappa}$  .
Using these values we get,
\begin{equation}
\psi_1^{\prime} = \frac{(\omega_1 + 2\sqrt{\alpha^2 - \omega_2^2})\nu_1 \sin^2 \theta }{\nu_1^2 \sin^2\theta - 2\omega_1 \sqrt{\alpha^2 - \omega_2^2} \cos^2 \theta + \omega_2^2 - \alpha^2} \ ,
\end{equation}
\begin{equation}
\psi_2^{\prime} = 0 \ ,
\end{equation}
and
\begin{equation}
\theta^{\prime} = \frac{\sqrt{3(\omega_1^2-\nu_1^2)(\alpha^2-\omega_2^2)} \cos \theta \sin \theta
\sqrt{\sin^2\theta - \sin^2 \theta_0}}{\nu_1^2 \sin^2\theta - 2\omega_1 \sqrt{\alpha^2 - \omega_2^2} \cos^2 \theta + \omega_2^2 - \alpha^2} \ ,
\end{equation}
where $$\sin^2 \theta_0 = \frac{\alpha^2 + 4\omega_1^2- \omega_2^2 + 4\omega_1\sqrt{\alpha^2 - \omega_2^2}}{3(\omega_1^2 - \nu_1^2)} \ .$$
%Note that if we put $\omega_2=0$ and replace $\omega_1$ by $\nu_2$ then these expressions will exactly %become that we obtained in our earlier paper(i.e. 1103.6153).
Now we can calculate the conserved quantities as,
\begin{equation}
E= -2T\int_{\theta_0}^{\frac{\pi}{2}} \frac{d\theta}{\theta^{\prime}}\frac{\partial \mathcal{L}}{\partial
\dot{t}} = \frac{2\kappa T (\alpha^2 - \nu_1^2 - \omega_2^2)}{(\alpha^2 - \omega_2^2) \sqrt{3(\omega_1^2 -
\nu_1^2)}} \int_{\theta_0}^{\frac{\pi}{2}} \frac{d\theta \sin\theta}{\cos\theta \sqrt{\sin^2\theta -
\sin^2\theta_0}} \ ,
\end{equation}
\begin{equation}
P_y= 2T\int_{\theta_0}^{\frac{\pi}{2}} \frac{d\theta}{\theta^{\prime}}\frac{\partial \mathcal{L}}{\partial
\dot{y}} = \frac{2 v T (\alpha^2 - \nu_1^2 - \omega_2^2)}{(\alpha^2 - \omega_2^2) \sqrt{3(\omega_1^2 -
\nu_1^2)}} \int_{\theta_0}^{\frac{\pi}{2}} \frac{d\theta \sin\theta}{\cos\theta \sqrt{\sin^2\theta -
\sin^2\theta_0}}  \ ,
\end{equation}
\begin{equation}
D_1= 2T\int_{\theta_0}^{\frac{\pi}{2}} \frac{d\theta}{\theta^{\prime}}\frac{\partial \mathcal{L}}{\partial
\dot{\rho}_1} = \frac{2 m_1 T (\alpha^2 - \nu_1^2 - \omega_2^2)}{(\alpha^2 - \omega_2^2) \sqrt{3(\omega_1^2 - \nu_1^2)}} \int_{\theta_0}^{\frac{\pi}{2}} \frac{d\theta \sin\theta}{\cos\theta \sqrt{\sin^2\theta -
\sin^2\theta_0}} \ ,
\end{equation}
\begin{equation}
D_2= 2T\int_{\theta_0}^{\frac{\pi}{2}} \frac{d\theta}{\theta^{\prime}}\frac{\partial \mathcal{L}}{\partial  \dot{\rho}_2} = \frac{2 m_2 T (\alpha^2 - \nu_1^2 - \omega_2^2)}{(\alpha^2 - \omega_2^2) \sqrt{3(\omega_1^2 - \nu_1^2)}} \int_{\theta_0}^{\frac{\pi}{2}} \frac{d\theta \sin\theta}{\cos\theta \sqrt{\sin^2\theta -
\sin^2\theta_0}} \ ,
\end{equation}
and finally the angular momentum $K_2$ is given by,
\begin{equation}
K_2= 2T\int_{\theta_0}^{\frac{\pi}{2}} \frac{d\theta}{\theta^{\prime}}\frac{\partial \mathcal{L}}{\partial \dot{\psi}_2} = \frac{2 \omega_2 T (\alpha^2 - \nu_1^2 - \omega_2^2)}{(\alpha^2 - \omega_2^2) \sqrt{3(\omega_1^2 - \nu_1^2)}} \int_{\theta_0}^{\frac{\pi}{2}} \frac{d\theta \sin\theta}{\cos\theta \sqrt{\sin^2\theta - \sin^2\theta_0}} \ .
\end{equation}
They all diverge, their combination however can be written in as,
\begin{equation}
\sqrt{E^2 - P_y^2 - D_1^2 - D_2^2 -K_2^2} =  \frac{2 T (\alpha^2 - \nu_1^2 - \omega_2^2)}{\sqrt{3(\omega_1^2 - \nu_1^2)(\alpha^2 - \omega_2^2)}} \int_{\theta_0}^{\frac{\pi}{2}} \frac{d\theta \sin\theta}{\cos\theta
\sqrt{\sin^2\theta - \sin^2\theta_0}} \ .
\end{equation}
The so called deficit angle $\Delta \phi$ also diverges, but the quantity
\begin{equation}
\Delta \phi^{\prime} = \Delta \phi + \frac{1}{T} \sqrt{E^2 - P_y^2 - D_1^2 - D_2^2 - K_2^2} = 2\theta_0  -\pi
\end{equation}
is finite.
The regularised angular momenta $J_1$ and $K_1$ are given by,
\begin{eqnarray}
\tilde{J}_1 = J_1 - \frac{2\nu_1 (\omega_1 + 2\sqrt{\alpha^2 - \omega_2^2})}{(\alpha^2 - \nu_1^2 -  \omega_2^2)} \sqrt{E^2 - P_y^2 - D_1^2 - D_2^2 - K_2^2} \nonumber \\ = \frac{-2\nu_1 T (4\omega_1 + 5\sqrt{\alpha^2 - \omega_2^2})}{\sqrt{3(\alpha^2 - \omega_2^2)(\omega_1^2 - \nu_1^2)}} \cos\theta_0  \ ,
\end{eqnarray}
\begin{eqnarray}
\tilde{K}_1 = K_1 - 2 \sqrt{E^2 - P_y^2 - D_1^2 - D_2^2 - K_2^2} = \frac{ 2T (4\nu_1^2 + 5\omega_1 \sqrt{\alpha^2 - \omega_2^2})}{\sqrt{3(\alpha^2 - \omega_2^2)(\omega_1^2 - \nu_1^2)}} \cos\theta_0 \ .
\nonumber \\
\end{eqnarray}
If we define $\tilde{J_1}=-\tilde{J}$, then it is easy to show that the angular momenta $\tilde{J}$ and $\tilde{K}_1$ satisfy the following
relation (for the parameters $\omega_2 = 0, \alpha = \nu_1$ ),
\begin{eqnarray}
\tilde{J} =
%\sqrt{\tilde{K}_1 - \frac{(25\alpha^2 - 16\nu_1^2 -25\omega_2^2)}{3(\alpha^2 - \omega_2^2)} %\frac{\lambda}{\pi^2} \sin^2(\frac{\pi}{2}-\theta_0)} \\ \nonumber &=& \sqrt{\tilde{K}_1 - %\frac{(25\alpha^2 - 16\nu_1^2)}{3\alpha^2} \frac{\lambda}{\pi^2} \sin^2(\frac{\pi}{2}-\theta_0)}
%|_{\omega_2 = 0} \\ \nonumber &=&
\sqrt{\tilde{K}_1 -  \frac{3\lambda}{\pi^2} \sin^2\left(\frac{\pi}{2}-\theta_0\right)} \ .
\end{eqnarray}
Note that this expression matches exactly with the ones derived in \cite{Biswas:2011wu}. We wish to stress
that in the present case the problem essentially reduces to that of rigidly rotating open string in the NS5-brane background.
\subsubsection{Giant Magnon solution}
In the opposite limiting case we get the condition $\alpha^2 = \nu_1^2 + \omega_1^2$, and by putting $C_4=2\nu_1$ and $C_5=-\frac{C_1\omega_2}{\kappa}$, we obtain the following
\begin{eqnarray}
\psi_1^{\prime} = -\frac{(C_1\omega_1 + 2\kappa \nu_1)\sin^2\theta}{(C_1\nu_1 +
2\kappa\omega_1)\cos^2\theta} \ , \>\>\> \psi_2^{\prime} = 0 \ ,
\end{eqnarray}
and
\begin{equation}
\theta^{\prime} = \frac{\sin \theta \sqrt{\sin^2 \theta - \sin^2 \theta_1}}{\sin\theta_1 \cos\theta} \ ,
\end{equation}
where $$\sin\theta_1 = \frac{C_1\nu_1 + 2\kappa\omega_1}{\kappa \sqrt{3(\omega_1^2-\nu_1^2)}} \ .$$
Now the conserved quantities become
\begin{equation}
E = -\frac{2T(\kappa^2 - C_1^2)}{\kappa \sqrt{3(\omega_1^2 - \nu_1^2)}} \int_{\theta_1}^{\frac{\pi}{2}}
\frac{d\theta \sin\theta}{\cos\theta \sqrt{\sin^2\theta - \sin^2 \theta_1}} \ ,
\end{equation}
\begin{equation}
P_y = -\frac{2vT(\kappa^2 - C_1^2)}{\kappa^2 \sqrt{3(\omega_1^2 - \nu_1^2)}} \int_{\theta_1}^{\frac{\pi}{2}} \frac{d\theta \sin\theta}{\cos\theta \sqrt{\sin^2\theta - \sin^2 \theta_1}} \ ,
\end{equation}
\begin{equation}
D_1 = -\frac{2m_1T(\kappa^2 - C_1^2)}{\kappa^2 \sqrt{3(\omega_1^2 - \nu_1^2)}}
\int_{\theta_1}^{\frac{\pi}{2}} \frac{d\theta \sin\theta}{\cos\theta \sqrt{\sin^2\theta - \sin^2 \theta_1}} \ ,
\end{equation}
\begin{equation}
D_2 = -\frac{2m_2T(\kappa^2 - C_1^2)}{\kappa^2 \sqrt{3(\omega_1^2 - \nu_1^2)}} \int_{\theta_1}^{\frac{\pi}{2}} \frac{d\theta \sin\theta}{\cos\theta \sqrt{\sin^2\theta - \sin^2 \theta_1}} \ ,
\end{equation}
\begin{equation}
K_2 = -\frac{2\omega_2T(\kappa^2 - C_1^2)}{\kappa^2 \sqrt{3(\omega_1^2 - \nu_1^2)}} \int_{\theta_1}^{\frac{\pi}{2}} \frac{d\theta \sin\theta}{\cos\theta \sqrt{\sin^2\theta - \sin^2 \theta_1}} \ .
\end{equation}
All these quantities diverges and their combination is given by,
\begin{equation}
\sqrt{E^2 - P_y^2 - D_1^2 - D_2^2 - K_2^2} = \frac{2\sqrt{\alpha^2 - \omega_2^2} T(\kappa^2 - C_1^2)}{\kappa^2 \sqrt{3(\omega_1^2 - \nu_1^2)}} \int_{\theta_1}^{\frac{\pi}{2}} \frac{d\theta \sin\theta}{\cos\theta \sqrt{\sin^2\theta - \sin^2 \theta_1}} \ .
\end{equation}
The deficit angle
\begin{equation}
\Delta \phi = 2T\int_{\theta_1}^{\frac{\pi}{2}}\frac{d\theta}{\theta^{\prime}} = \pi - 2\theta_1
\end{equation}
is however finite. The regularised $J_1$ is given by,
\begin{eqnarray}
\tilde{J}_1 &=& J_1 + \frac{(5\kappa^2\nu_1 + 2C_1\kappa \omega_1 - C_1^2\nu_1)}{\sqrt{\alpha^2 -
\omega_2^2}(\kappa^2 - C_1^2)} \sqrt{E^2 - P_y^2 - D_1^2 - D_2^2 - K_2^2} \nonumber \\ &=&
\frac{2T(4C_1\omega_1 + 5\kappa \nu_1)}{\kappa \sqrt{3(\omega_1^2 - \nu_1^2)}} \cos\theta_1 \ .
\end{eqnarray}
Angular momenta $K_1$ is once again finite and is given by,
\begin{equation}
K_1 = -\frac{2T(4C_1\nu_1 + 5\kappa \omega_1)}{\kappa \sqrt{3(\omega_1^2 - \nu_1^2)}} \cos\theta_1
\end{equation}
To know a precise relation between the energy, angular momentum and the distance between two end point
of the string, we rescale the expression $\sqrt{E^2 - P_y^2 - D_1^2 - D_2^2 - K_2^2}$ as follows,
\begin{equation}
\tilde{E} = \frac{(5\kappa^2\nu_1 + 2C_1\kappa \omega_1 - C_1^2\nu_1)}{\sqrt{\alpha^2 - \omega_2^2}(\kappa^2 - C_1^2)} \sqrt{E^2 - P_y^2 - D_1^2 - D_2^2 - K_2^2} \ .
\end{equation}
If we define $J=-J_1$,  then by looking at the form of various quantities, it is easy to derive (for $C_1 = \kappa$)
\begin{eqnarray}
\tilde{E} -  J
%&=& \sqrt{K_1^2 - \frac{(25\kappa^2 - 16C_1^2)}{3\kappa^2} \frac{\lambda}{\pi^2} \sin^2 \frac{\Delta %\phi}{2}} \\ \nonumber
= \sqrt{K_1^2 -  \frac{3\lambda}{\pi^2} \sin^2 \frac{\Delta \phi}{2}} \ .
\end{eqnarray}
Once again, this relationship was derived in \cite{Biswas:2011wu}, where $\Delta \phi$ was identified as the
worldsheet momentum along the string worldsheet.

\subsection{{\rm For} $\theta_1 = \theta_2 = \theta$} In this subsection we wish to
 study the solution of the rotating string when it rotates around both the spheres. To make our calculations
 doable and simple, we will restrict ourselves to the case when $\theta_1 = \theta_2 = \theta$.
To this end, let us define
\begin{eqnarray}
a=C_1(\nu_1^2 + \nu_2^2)\sin^2\theta - C_1 \alpha^2,\>\>\>
b=C_1\omega_2^2 \cos^2\theta + 2\kappa \nu_1 \omega_1 \sin^2\theta - \kappa \omega_1 C_4, \nonumber \\
c=C_1\omega_1^2 \cos^2\theta + 2\kappa \nu_2 \omega_2 \sin^2 \theta - \kappa \omega_2 C_5, \>\>\>
d=(\nu_1+\nu_2)\sin^2 \theta , \nonumber \\
e= \kappa C_4 -2\kappa \nu_1\sin^2\theta + C_1 \omega_1\cos^2 \theta, \>\>\>
f= \kappa C_5 -2\kappa \nu_2 \sin^2 \theta + C_1 \omega_2\cos^2 \theta . \nonumber \\
\end{eqnarray}
With the above identifications, first two equations (\ref{1st}) and (\ref{2nd}) reduce to
\begin{eqnarray}
(a + b)\psi_1^{\prime}\cos^2 \theta &=& e(\omega_2 \psi_2^{\prime}\cos^2 \theta + d) \ , \nonumber \\
(a + c)\psi_2^{\prime}\cos^2 \theta &=& f(\omega_1 \psi_1^{\prime}\cos^2 \theta + d) \ .
\end{eqnarray}
From these we can easily solve for $\psi_1^{\prime}$ and $\psi_2^{\prime}$, and they are given by,
\begin{equation}
\psi_1^{\prime} = \frac{de(a + c + \omega_2 f)}{\cos^2\theta[(a+b)(a+c) - \omega_1 \omega_2 ef]} \ ,
\end{equation}
\begin{equation}
\psi_2^{\prime} = \frac{df(a + b + \omega_1 e)}{\cos^2\theta[(a+b)(a+c) - \omega_1 \omega_2 ef]} \ .
\end{equation}
Further (\ref{theta}) is now given by
\begin{equation}
{\theta^{\prime}}^2 = \frac{\kappa^2 - C_1^2}{2 C_1^2} \frac{[(\nu_1 +\nu_2) \sin^2 \theta
+ (\omega_1 \psi_1^{\prime}+  \omega_2 \psi_2^{\prime}) \cos^2\theta]^2}{[\alpha^2 - (\nu_1^2
+ \nu_2^2) \sin^2\theta - (\omega_1^2 + \omega_2^2) \cos^2\theta]} - \sin^2 \theta
- \frac{1}{2}\left({\psi_1^{\prime}}^2 + {\psi_2^{\prime}}^2\right) \cos^2 \theta
\end{equation}
Below we will discuss the different limits that correspond to the single spike and giant
magnon solutions.
\subsubsection{Single Spike solution}
In the limit $\theta^{\prime} \rightarrow 0$ as $\theta \rightarrow \frac{\pi}{2}$, we obtain the
condition
\begin{eqnarray}
C_1= \frac{\kappa (\nu_1 + \nu_2)}{\alpha_1} \ ,
\end{eqnarray}
where $\alpha_1= \sqrt{2\alpha^2 - (\nu_1 - \nu_2)^2} $ and
$\psi_1^{\prime} \rightarrow 0$ and $\psi_2^{\prime} \rightarrow 0$ as $\theta \rightarrow
\frac{\pi}{2}$ implies $C_4=2\nu_1$ and $C_5=2\nu_2$ respectively. Using these values we get,
\begin{eqnarray}
\psi_1^{\prime} = \frac{(C_1 \omega_1 + 2\kappa \nu_1)(\nu_1 + \nu_2) \sin^2\theta}{C_1(\nu_1^2 +
\nu_2^2)\sin^2\theta - C_1 \alpha^2 - 2\kappa \alpha_2 \cos^2 \theta} \ ,
\end{eqnarray}
and
\begin{equation}
\psi_2^{\prime} = \frac{(C_1 \omega_2 + 2\kappa \nu_2)(\nu_1 + \nu_2) \sin^2\theta}{C_1(\nu_1^2 +
\nu_2^2)\sin^2\theta - C_1 \alpha^2 - 2\kappa \alpha_2 \cos^2 \theta} \ .
\end{equation}
where $\alpha_2 = \nu_1 \omega_1 + \nu_2 \omega_2 $.
Now the equation for $\theta$ can be written down as
\begin{equation}
\theta^{\prime} = \frac{\sqrt{\frac{h}{2}} \sin \theta \cos \theta \sqrt{\sin^2 \theta - \sin^2
\theta_0}}{(\nu_1 + \nu_2)(\nu_1^2 + \nu_2^2)\sin^2 \theta - 2\alpha_1 \alpha_2 \cos^2 \theta - (\nu_1 + \nu_2)\alpha^2} \ ,
\end{equation}
where
\begin{eqnarray}
\sin^2\theta_0 = \frac{2}{h}\left[(\nu_1 + \nu_2)\alpha^2 + 2\alpha_1 \alpha_2 \right]^2 \ ,
\end{eqnarray}
with
\begin{eqnarray}
h = (\nu_1 + \nu_2)^2 [(\nu_1^2 + \nu_2^2)(2\alpha^2 - 4\alpha_1^2) -\alpha_1^2(\omega_1^2 + \omega_2^2)]
+ 8 \alpha_1^2 \alpha_2^2
+4 \alpha_1 \alpha_2 (\nu_1 + \nu_2) (\nu_1 -
\nu_2)^2  \ . \nonumber \\
\end{eqnarray}
Now we can calculate the conserved quantities as
\begin{equation}
E = -2T \int_{\theta_0}^{\frac{\pi}{2}} \frac{d\theta}{\theta^{\prime}} \frac{\partial
\mathcal{L}}{\partial \dot{t}} = \frac{4\kappa T (\nu_1+\nu_2)(\alpha^2-\nu_1^2-\nu_2^2)}{\sqrt{\frac{h}{2}}
\alpha_1} \int_{\theta_0}^{\frac{\pi}{2}} \frac{d\theta \sin \theta}{\cos \theta\sqrt{\sin^2
\theta -\sin ^2 \theta_0}} \ ,\nonumber \\
\end{equation}
\begin{equation}
P_y = 2T \int_{\theta_0}^{\frac{\pi}{2}} \frac{d\theta}{\theta^{\prime}} \frac{\partial \mathcal{L}}{\partial \dot{y}} = \frac{4 v T (\nu_1+\nu_2)(\alpha^2-\nu_1^2-\nu_2^2)}{\sqrt{\frac{h}{2}} \alpha_1}
\int_{\theta_0}^{\frac{\pi}{2}} \frac{d\theta \sin \theta}{\cos \theta\sqrt{\sin^2 \theta -\sin ^2 \theta_0}} , \nonumber \\
\end{equation}
\begin{equation}
D_1 = 2T \int_{\theta_0}^{\frac{\pi}{2}} \frac{d\theta}{\theta^{\prime}} \frac{\partial \mathcal{L}}{\partial
\dot{\rho}_1} = \frac{4 m_1 T (\nu_1+\nu_2)(\alpha^2-\nu_1^2-\nu_2^2)}{\sqrt{\frac{h}{2}}
\alpha_1} \int_{\theta_0}^{\frac{\pi}{2}} \frac{d\theta \sin \theta}{\cos \theta\sqrt{\sin^2
\theta -\sin ^2 \theta_0} } , \nonumber \\
\end{equation}
\begin{equation}
D_2 = 2T \int_{\theta_0}^{\frac{\pi}{2}} \frac{d\theta}{\theta^{\prime}} \frac{\partial \mathcal{L}}{\partial
\dot{\rho}_2} = \frac{4 m_2 T (\nu_1+\nu_2)(\alpha^2-\nu_1^2-\nu_2^2)}{\sqrt{\frac{h}{2}}
\alpha_1} \int_{\theta_0}^{\frac{\pi}{2}} \frac{d\theta \sin \theta}{\cos \theta\sqrt{\sin^2
\theta -\sin ^2 \theta_0} } .\nonumber \\
\end{equation}
All the conserved quantities $E$, $P_y$, $D_1$ and $D_2$ described above are infinite. For our later use,
we define the following rescaled quantity,
\begin{equation}
\sqrt{E^2-P_y^2-D_1^2-D_2^2} = \frac{4\alpha T (\nu_1+\nu_2)(\alpha^2-\nu_1^2-\nu_2^2)}{\sqrt{\frac{h}{2}}
\alpha_1} \int_{\theta_0}^{\frac{\pi}{2}} \frac{d\theta \sin \theta}{\cos \theta\sqrt{\sin^2
\theta -\sin ^2 \theta_0} } \ .
\end{equation}
Now the Deficit angle is defined by $\Delta \phi= 2 \int_{\theta_0}^{\frac{\pi}{2}} \frac{d\theta}{\theta^{\prime}}$ is infinite but
\begin{eqnarray}
\Delta \phi^{\prime} = \Delta \phi + \frac{\alpha_1}{2\alpha T}\sqrt{E^2-P_y^2-D_1^2-D_2^2} = 2\theta_0 -\pi \nonumber \\
\end{eqnarray}
is finite. Similarly the regularized $J_1$, $J_2$, $K_1$, and $K_2$ are given by
\begin{eqnarray}
\tilde{J_1} &=& J_1 - \frac{[(\nu_1 - \nu_2)(\alpha^2 - \nu_1^2 - \nu_2^2) + 2\alpha_1 \omega_1 (\nu_1+\nu_2) + 4 \alpha_1^2\nu_1]}{2\alpha (\alpha^2-\nu_1^2-\nu_2^2)}
\sqrt{E^2-P_y^2-D_1^2-D_2^2} \nonumber \\ &=& -\frac{2T (\nu_1+\nu_2) \cos
\theta_0}{\alpha_1 \sqrt{\frac{h}{2}}}\left[ (\nu_2 - \nu_1)(\nu_1^2 + \nu_2^2) + 2\alpha_1 \omega_1 (\nu_1 + \nu_2) + 2\nu_1(\alpha^2 + 2\alpha_1^2) + 2\alpha_1\alpha_2 \right] , \nonumber \\
\end{eqnarray}
\begin{eqnarray}
\tilde{J_2} &=& J_2 - \frac{[(\nu_2 - \nu_1)(\alpha^2 - \nu_1^2 - \nu_2^2) + 2\alpha_1 \omega_2 (\nu_1+\nu_2) + 4 \alpha_1^2\nu_2]}{2\alpha (\alpha^2-\nu_1^2-\nu_2^2)}
\sqrt{E^2-P_y^2-D_1^2-D_2^2} \nonumber \\ &=& -\frac{2T (\nu_1+\nu_2) \cos
\theta_0}{\alpha_1 \sqrt{\frac{h}{2}}}\left[ (\nu_1 - \nu_2)(\nu_1^2 + \nu_2^2) + 2\alpha_1 \omega_2 (\nu_1 + \nu_2) + 2\nu_2(\alpha^2 + 2\alpha_1^2) + 2\alpha_1\alpha_2 \right] , \nonumber \\
\end{eqnarray}
\begin{eqnarray}
\tilde{K}_1 = \frac{2T \cos \theta_0}{\sqrt{\frac{h}{2}}}\left[ (\nu_1+ \nu_2)(\alpha_1 \omega_1 + 2(2\nu_1^2 + \nu_1\nu_2  + \nu_2^2 )) + 4\alpha_1\alpha_2 \right] \ ,
\end{eqnarray}
\begin{eqnarray}
\tilde{K}_2 = \frac{2T \cos \theta_0}{\sqrt{\frac{h}{2}}}\left[ (\nu_1+ \nu_2)(\alpha_1 \omega_2 + 2(\nu_1^2 + \nu_1\nu_2  + 2\nu_2^2 )) + 4\alpha_1\alpha_2 \right]
\end{eqnarray}
where $\tilde{K}_i = K_i -\frac{\alpha_1}{\alpha}\sqrt{E^2-P_y^2-D_1^2-D_2^2}$, with $i= 1, 2$.
Now if define $\tilde{J} = \tilde{J_1} - \tilde{J_2}$ and $\tilde{K}= \tilde{K}_1 - \tilde{K}_2$ then we obtain,
%\begin{eqnarray}
%\tilde{J} = \frac{4T(\nu_1+\nu_2)\cos \theta_0}{\alpha_1 \sqrt{\frac{h}{2}}} \left[(\nu_2-\nu_1)(\alpha^2 - %\nu_1^2 - \nu_2^2 + 2\alpha_1^2) +
%\alpha_1(\omega_2-\omega_1)(\nu_1 +\nu_2)\right]
%\end{eqnarray}
% and
%\begin{eqnarray}
%\tilde{K} = \frac{2T(\nu_1+\nu_2)\cos \theta_0}{\sqrt{\frac{h}{2}}} [(\omega_1 - \omega_2)
%\alpha_1 + 2(\nu_1^2 - \nu_2^2)]
%\end{eqnarray}
%Note that although $K_1$ and $K_2$ diverges, but their difference $K=K_1-K_2$ is finite.
%These newly defined $\tilde{J}$ and $\tilde{K}$ satisfy the relation,
\begin{eqnarray}
\tilde{J} = \sqrt{\tilde{K}^2 + \frac{2g}{h}\frac{(\nu_1+\nu_2)^2}{\alpha_1^2} \frac{\lambda}{\pi^2} \sin^2(\frac{\pi}{2}- \theta_0)} \ ,
\label{dispersion-sp}
\end{eqnarray}
where
\begin{eqnarray}
g = \left[2(\nu_2-\nu_1)(\alpha^2 - \nu_1^2 - \nu_2^2 + 2\alpha_1^2) -
2\alpha_1(\omega_1-\omega_2)(\nu_1 +\nu_2) \right]^2 \nonumber \\
-  \left[ 2\alpha_1 (\nu_1^2 - \nu_2^2)  + \alpha_1^2(\omega_1 - \omega_2) \right]^2 \nonumber \\
h = (\nu_1 + \nu_2)^2 [(\nu_1^2 + \nu_2^2)(2\alpha^2 - 4\alpha_1^2) -\alpha_1^2(\omega_1^2 + \omega_2^2)]
+ 8 \alpha_1^2 \alpha_2^2 \nonumber \\
+4 \alpha_1 \alpha_2 (\nu_1 + \nu_2) (\nu_1 - \nu_2)^2 \ .
\end{eqnarray}
Few comments are in order. The dispersion relation obtained has complicated prefactor multiplied with
$\frac{\lambda}{\pi^2}$ can
be attributed to the fact that this dispersion relation is not of a single spike when the string
rigidly rotates around the sphere, but is supported by a large number of other charges including that of
NS-NS B-field which supports the background and the fundamental string knows the presence of such
charges through the open string boundary condition. Furthermore the presence of charges like $D_1$
and $D_2$ do not imply any new interpretation of the dispersion relation. It simply reflects the
fact that the motion of the string in the radial direction in the near horizon geometry of I-brane is free.
\subsubsection{Giant Magnon solution}
In the opposite limit corresponding to the one just described above, we get
$$\alpha^2 = \nu_1^2 + \nu_2^2 \ ,$$
and by putting  $C_4=2\nu_1$ and $C_5=2\nu_2$ we get
\begin{equation}
\psi_1^{\prime} = -\frac{(\nu_1 + \nu_2)(C_1 \omega_1 + 2\kappa \nu_1) \sin^2 \theta}{\beta \cos^2 \theta} \ ,
\end{equation}
\begin{equation}
\psi_2^{\prime} = -\frac{(\nu_1 + \nu_2)(C_1 \omega_2 + 2\kappa \nu_2) \sin^2 \theta}{\beta \cos^2 \theta} \ ,
\end{equation}
where $\beta = 2\kappa \nu_1 \omega_1 + 2\kappa \nu_2 \omega_2 + C_1 \alpha^2$ and
\begin{equation}
\theta^{\prime} = \frac{\sin \theta \sqrt{\sin^2 \theta - \sin^2 \theta_1}}{\sin \theta_1 \cos \theta} \ ,
\end{equation}
where $\sin^2 \theta_1= \frac{2\beta^2}{h_1}$
and
\begin{eqnarray}
h_1=(\nu_1 + \nu_2)^2[(\kappa^2 - C_1^2)(\alpha^2 - \omega_1^2 - \omega_2^2) - (C_1\omega_1 + 2\kappa \nu_1)^2 - (C_1\omega_2 + 2\kappa \nu_2)^2] + 2\beta^2  \ . \nonumber \\
\end{eqnarray}
The conserved quantities are
\begin{equation}
E= -\frac{2T (\nu_1+\nu_2)(\kappa^2 - C_1^2)}{\sqrt{\frac{h_1}{2}} } \int_{\theta_1}^{\frac{\pi}{2}}
\frac{d\theta \sin \theta}{\cos \theta\sqrt{\sin^2 \theta -\sin ^2 \theta_1}} \ ,
\end{equation}
\begin{equation}
P_y = -\frac{2 v T (\nu_1+\nu_2)(\kappa^2 - C_1^2)}{\kappa \sqrt{\frac{h_1}{2}} }
\int_{\theta_1}^{\frac{\pi}{2}} \frac{d\theta \sin \theta}{\cos \theta\sqrt{\sin^2 \theta -\sin ^2 \theta_1}} \ ,
\end{equation}
\begin{equation}
D_1 = -\frac{2 m_1 T (\nu_1+\nu_2)(\kappa^2 - C_1^2)}{\kappa \sqrt{\frac{h_1}{2}} }
\int_{\theta_1}^{\frac{\pi}{2}} \frac{d\theta \sin \theta}{\cos \theta\sqrt{\sin^2 \theta -\sin ^2 \theta_1} } \ ,
\end{equation}
\begin{equation}
D_2 = -\frac{2 m_2 T (\nu_1+\nu_2)(\kappa^2 - C_1^2)}{\kappa \sqrt{\frac{h_1}{2}} }
\int_{\theta_1}^{\frac{\pi}{2}} \frac{d\theta \sin \theta}{\cos \theta\sqrt{\sin^2 \theta -\sin ^2 \theta_1} } \ .
\end{equation}
Here also the conserved quantities $E$, $P_y$, $D_1$ and $D_2$ are all infinite and the rescaled
combination can be written as,
\begin{equation}
\sqrt{E^2-P_y^2-D_1^2-D_2^2} = \frac{2\alpha T (\nu_1+\nu_2)(\kappa^2 - C_1^2)}{\kappa \sqrt{\frac{h_1}{2}} } \int_{\theta_1}^{\frac{\pi}{2}} \frac{d\theta \sin \theta}{\cos \theta\sqrt{\sin^2 \theta -\sin ^2
\theta_1}} \ .
\end{equation}
The deficit angle $\Delta \phi$ is however finite and is given by
\begin{equation}
\Delta \phi = 2\int_{\theta_1}^{\frac{\pi}{2}} \frac{d\theta}{\theta^{\prime}} =2\int_{\theta_1}^{\frac{\pi}{2}} \frac{d\theta \sin\theta_1 \cos\theta}{\sin\theta \sqrt{\sin^2\theta-\sin^2\theta_1}} = \pi-2\theta_1 \ .
\end{equation}
The regularized $J_1$ and $J_2$ are given by,
\begin{eqnarray}
&&\tilde{J_1} = J_1 + \frac{(5\kappa^2\nu_1 + 2C_1\kappa \omega_1 - C_1^2\nu_1)}{\alpha(\kappa^2 - C_1^2)} \sqrt{E^2-P_y^2-D_1^2-D_2^2} \nonumber \\ &=& \frac{2T \cos \theta_1}{\kappa \sqrt{\frac{h_1}{2}}}[C_1\beta +  (\nu_1 + \nu_2)(5\kappa^2\nu_1 + 2C_1\kappa \omega_1 - C_1^2\nu_1) ]  \ ,\nonumber \\
\end{eqnarray}
and
\begin{eqnarray}
&&\tilde{J_2} = J_2 + \frac{(5\kappa^2\nu_2 + 2C_1\kappa \omega_2 - C_1^2\nu_2)}{\alpha(\kappa^2 - C_1^2)} \sqrt{E^2-P_y^2-D_1^2-D_2^2} \nonumber \\ &=& \frac{2T \cos \theta_1}{\kappa \sqrt{\frac{h_1}{2}}}[C_1\beta +  (\nu_1 + \nu_2)(5\kappa^2\nu_2 + 2C_1\kappa \omega_2 - C_1^2\nu_2) ] \ , \nonumber \\
\end{eqnarray}
The other two angular momenta are finite and are given by,
\begin{eqnarray}
K_1 &=& -\frac{2T \cos \theta_1}{ \sqrt{\frac{h_1}{2}}}[(\nu_1 + \nu_2)(\kappa \omega_1 + 2C_1 \nu_1)
+  2\beta]  ,
\end{eqnarray}
and
\begin{eqnarray}
K_2&=& -\frac{2T \cos \theta_1}{\sqrt{\frac{h_1}{2}}}[(\nu_1 + \nu_2)(\kappa \omega_2 + 2C_1 \nu_2)
+ 2\beta]  ,
\end{eqnarray}
Similarly as before, if we define $\tilde{J}=\tilde{J_1} - \tilde{J_2}$ , $K= K_1 - K_2$ and $J=J_2-J_1$ then we obtain,
%\begin{equation}
%\tilde{J} = \frac{2T(\nu_1+\nu_2)\cos \theta_1}{\kappa \sqrt{\frac{h_1}{2}}} [2\kappa C_1 (\omega_1 - %\omega_2) + (5 \kappa^2 - C_1^2)(\nu_1 - \nu_2)]
%\end{equation}
%and
%\begin{equation}
%K = -\frac{2T(\nu_1+\nu_2)\cos \theta_1}{\kappa \sqrt{\frac{h_1}{2}}} [\kappa^2 (\omega_1 - \omega_2) + %2C_1 \kappa (\nu_1-\nu_2)]
%\end{equation}
%Again, if we define $\tilde{E} = \frac{2\kappa C_1 (\omega_1 - \omega_2) + (5 \kappa^2 - C_1^2)(\nu_1 - %\nu_2)}{\alpha(\kappa^2 - C_1^2)} \sqrt{E^2-P_y^2-D_1^2-D_2^2}$ then it satisfy the relation,

\begin{equation}
\tilde{E} - J = \sqrt{K^2 + \frac{2g_1}{h_1} \frac{(\nu_1+\nu_2)^2}{\kappa^2}
\frac{\lambda}{\pi^2} \sin^2 \frac{\Delta \phi}{2}} \ ,
\end{equation}
where
\begin{eqnarray}
\tilde{E} = \frac{2\kappa C_1 (\omega_1 - \omega_2) + (5 \kappa^2 - C_1^2)(\nu_1 - \nu_2)}{\alpha(\kappa^2 - C_1^2)} \sqrt{E^2-P_y^2-D_1^2-D_2^2} \ , \nonumber \\
g_1=[2\kappa C_1 (\omega_1 - \omega_2) + (5 \kappa^2 - C_1^2)(\nu_1 - \nu_2)]^2 - [\kappa^2 (\omega_1 - \omega_2) + 2C_1 \kappa (\nu_1-\nu_2)]^2 , \nonumber \\
h_1=(\nu_1 + \nu_2)^2[(\kappa^2 - C_1^2)(\alpha^2 - \omega_1^2 - \omega_2^2) - (C_1\omega_1 + 2\kappa \nu_1)^2 - (C_1\omega_2 + 2\kappa \nu_2)^2] + 2\beta^2 \ .
\nonumber \\
\end{eqnarray}
Once again we would like to stress that this dispersion relation must be compared with that
in the presence of other background fields. It will be really a challenge to find the dual states
in the corresponding field theory side.
\section{Conclusions}
We have studied in this paper the rigidly rotating strings in the near horizon geometry of a
1+1 dimensional intersection of two stacks of orthogonal branes, the so called I-brane background.
We have studied the solutions of fundamental string action in this background in the presence of
various NS-NS B-fields. Because the fundamental string couple to the background NS-NS field, the
string dynamics knows the presence of these fluxes and the dispersion relation also reflects
this fact. As the background in question is too difficult to solve exactly we have assumed first the
motion of the string constrained only on one sphere and found out two solutions corresponding to
single spike and giant magnons. This procedure effectively means as if we are considering the
motion of the string in the near horizon geometry of one stack of NS5-branes.
Further we have generalized the solution to include motion on both
the spheres and write down corresponding dispersion relation for the giant magnon and single spike
solutions. We have also remarked that these solutions should be compared to the ones obtained in the
presence of other background fields that couple to the string. We have used the open string boundary
condition to explicitly observe the effect on the dispersion relation among various conserved charges.

There are various questions that can be persued further. As we know, the fact that
I-brane background is an exact solution of string theory equations of motion, helps us in
obtaining more information about the intersecting brane system itself which is otherwise not possible
via gauge theory. It will definitely be interesting to see from a probe brane analysis what more information
we get about the system which is apriori unknown from the field theory. Further it will be interesting to tell more
about the existence of some operators corresponding to the spiky strings from the point of view of dual theory,
whose exact nature in not known in detail. The supersymmetry of such states can be checked by following \cite{David:2008yk} in the bulk. Hence it will be interesting to see whether similar kind of states exist in the dual theory. It would also be interesting to whether such spiky strings exist as solutions in the worldvolume of
D-branes.
\vskip .2in
\noindent
{\bf Acknowledgements:} KLP would like to thank the Abdus Salam I.C.T.P, Trieste for hospitality under Associate
Scheme, where a part of this work was completed.

\end{document}